# Supply-Power-Constrained Cable Capacity Maximization Using Multi-Layer Neural Networks

Junho Cho, *Member, IEEE,* Sethumadhavan Chandrasekhar, *Fellow, IEEE,* Erixhen Sula, *Student Member, IEEE,* Samuel Olsson, Ellsworth Burrows, Greg Raybon, *Fellow, IEEE,* Roland Ryf, *Senior Member, IEEE,* Nicolas Fontaine, *Senior Member, IEEE,* Jean-Christophe Antona, *Member, IEEE,* Steve Grubb, *Member, IEEE,* Peter Winzer, *Fellow, IEEE,* and Andrew Chraplyvy, *Fellow, IEEE*

*Abstract*—We experimentally solve the problem of maximizing capacity under a total supply power constraint in a massively parallel submarine cable context, i.e., for a spatially uncoupled system in which fiber Kerr nonlinearity is not a dominant limitation. By using multi-layer neural networks trained with extensive measurement data acquired from a 12-span 744-km optical fiber link as an accurate digital twin of the true optical system, we experimentally maximize fiber capacity with respect to the transmit signal's spectral power distribution based on a gradient-descent algorithm. By observing convergence to approximately the same maximum capacity and power distribution for almost arbitrary initial conditions, we conjecture that the capacity surface is a concave function of the transmit signal power distribution. We then demonstrate that eliminating gain flattening filters (GFFs) from the optical amplifiers results in substantial capacity gains per Watt of electrical supply power compared to a conventional system that contains GFFs.

*Index Terms*—Optical fiber communication, channel capacity, artificial neural networks.

## I. INTRODUCTION

MASSIVE spatial parallelism has been shown to maximize the capacity and to minimize the cost/bit of submarine optical cables, which, in contrast to terrestrial transmission systems, are constrained by a fixed amount of electrical supply power per cable [1]–[4]: The capacity of a massively parallel system with a supply power constraint is given by [3], [5]

$$C_{Cable} = 2MB \log_2(1 + SNR)$$
$$= 2MB \log_2\left(1 + \eta \frac{P}{2MBN_0}\right), \quad (1)$$

where $SNR$ is the channel's *effective signal-to-noise ratio* (SNR), $\eta \leq 1$ is a constant that accounts for transponder implementation penalty, $P$ is the total signal power across all spatial paths, and $M$, $B$, $N_0$ denote the number of spatial paths, the system bandwidth, and the average noise power spectral density per spatial path, respectively. The channel's SNR is measured after coherent digital signal processing (DSP) at the input to the receiver's forward error correction (FEC) decoder. An alternative two-parameter version of Eq. (1) that has also been extensively verified experimentally is given by [6],

$$SNR = (\eta^{-1}SNR_{TRX}^{-1} + SNR_{Line}^{-1})^{-1}, \quad (2)$$

with the transponder's noise $N_{0,TRX}$ captured by an equivalent $SNR_{TRX}$ and the noise from the transmission line $N_{0,Line}$ captured by $SNR_{Line}$. The relations $SNR_{TRX} = \eta P/(2MBN_{0,TRX})$ and $SNR_{Line} = P/(2MBN_{0,Line})$ connect Eqns. (1) and (2), with $N_0 = N_{0,TRX} + N_{0,Line}$. If the transponder's noise contributions can be neglected due to the much larger noise contributions from the transmission line (i.e., if $\eta N_{0,Line} \gg N_{0,TRX}$), which is a legitimate assumption for an ultra-long-haul system, the effective SNR is simply given by $\eta\, SNR_{Line}$. For a uniform channel power distribution, $SNR_{Line}$ is the ratio of the overall received (RX) signal power $P$ to the overall RX noise power $2MBN_0$, with $N_0$ dominated by amplified spontaneous emission (ASE) [7] and nonlinear interference noise (NLIN) [8], [9] in the context of an auxiliary additive white Gaussian noise (AWGN) channel. Capacity of the channels with colored additive Gaussian noise is maximized by spectral waterfilling [10], [11, Ch. 9], although it is known that waterfilling only brings significant gains at very low SNRs [12, Ch. 5], [13], cf. Appendix A.

The cable capacity $C_{Cable}$ as per Eq. (1) *monotonically increases with* $M$, suggesting infinite spatial power dilution in

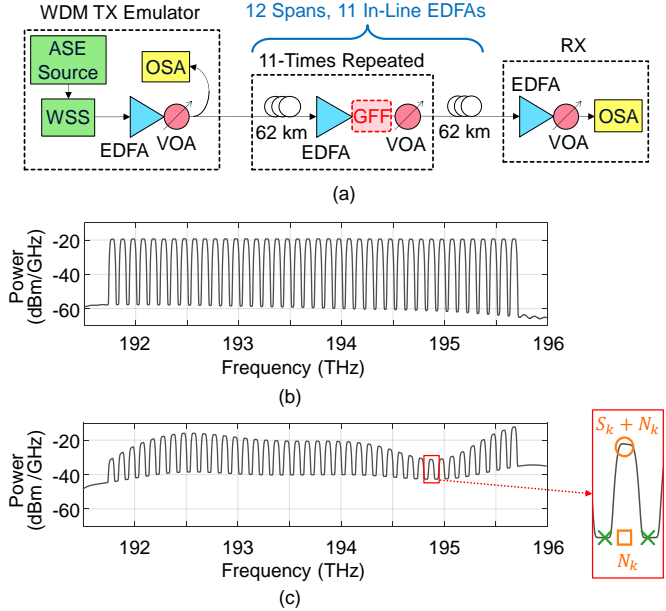

Fig. 1. (a) Experimental setup, (b) measured optical spectrum at the TX with a launch power of 13 dBm, (c) measured optical spectrum at the RX when all GFFs are removed from the link (*inset:* OSA-based SNR estimation).

ii) What is the optimum optical channel power allocation strategy in a waterfilling context in GFF-free transmission links, which are characterized by highly colored noise?

This paper addresses both topics. On an exemplary 12-span 744-km straight-line system, we experimentally achieve a capacity gain per Watt of electrical supply power of 19%. Higher gains in $m$ are expected for longer links and for pump-sharing architectures across amplifier arrays.

As we will show in Sec. II, accurately predicting the signal and noise power evolution of a long chain of un-flattened optical amplifiers for arbitrary transmit (TX) power profiles is difficult. A small change in the TX power spectral density (PSD) or in the spectral link characteristics may cause a complicated signal and noise power evolution through the system, making it intractable to computationally solve the problem using analytical or numerical physics-based optical amplifier models. We therefore resort to machine learning [17], [18] and build a multi-layer neural network (NN) as a *digital twin* of our optical fiber link. Once properly trained with experimental link data, the NN allows for an off-line gradient-descent (GD) optimization whose optimized results are then verified experimentally.

## II. EXPERIMENTAL METHODOLOGY AND SETUP

As discussed above, massively parallel submarine cables will operate at low-enough optical signal powers to neglect fiber nonlinearities. In addition, probabilistic constellation shaping allows to finely adapt each wavelength channel's transponder to the specific SNR of that channel [19]–[22]. This lets measurements of the delivered OSNR be a good basis for estimating polarization- and wavelength-division multiplexed (WDM) system capacities as

$$C = 2R_s \sum_{k=1}^{K} \log_2(1 + \eta\, SNR_k), \quad (4)$$

where $SNR_k$ (used for notational simplicity instead of $SNR_{Line,k}$ throughout the rest of this paper) is the OSNR of the $k$-th of $K$ WDM channels (normalized to one polarization and a reference bandwidth equal to the symbol rate $R_s$). Throughout this paper, we use $\eta = 1$ without limiting the generality of the optimization methodology (a greater gain from the proposed approach is expected with $\eta < 1$, as will be quantified at the end of Sec. IV.)

In order to determine $SNR_k$, we use the experimental WDM channel emulation method shown in Fig. 1: ASE from three serially concatenated erbium-doped fiber amplifiers (EDFAs) as an ASE source is filtered by a wavelength selective switch (WSS) to produce 40 ASE-loaded frequency slots of 50-GHz bandwidth that emulate 40 signal channels, a technique that is customarily used for WDM loading channels [23], [24]. These "signal" slots are interleaved with 39 "empty" 50-GHz slots across a 4-THz C-band system bandwidth, cf. Fig. 1(b). As shown in Fig. 1(c), $SNR_k$ at the receiver can then be estimated by an optical spectrum analyzer (OSA) taking the ratio of the emulated signal power $S_k$ (cf. circle marker in the inset, order to achieve the highest possible cable capacity. In systems with constant-output power amplifiers, Eq. (1) can be refined such that the signal power per spatial path decreases at a higher rate than $M$ as $M$ increases, while noise power increases at a lower rate than $M$, known as the *signal and noise droop* phenomenon [14], which results in a finite (yet very large) number of spatial paths to maximize $C_{Cable}$. Electrical-to-optical power conversion efficiency is another aspect that can further restrict the optimal degree of spatial parallelism under a supply-power constraint [15]. Eventually, cost considerations mostly limit the optimal number of spatial paths in a typical submarine cable to ~50-100 (~25-50 fiber pairs) [3]. Importantly, in all cases, the optical power dilution among many parallel fibers pushes transmission from nonlinearly-optimum launch powers to the linear transmission regime and lets $SNR_{Line}$ become exclusively the optical SNR (OSNR) due to the accumulation of in-line ASE. The logarithmically reduced spectral efficiency from a lower delivered OSNR per fiber is linearly over-compensated by the increased spatial multiplicity of the cable, cf. Eq. (1). In such systems, the capacity $C$ per Watt of electrical supply power $\mathcal{P}_E$, both per spatial path, becomes a key figure of merit [16]:

$$m = C/\mathcal{P}_E. \quad (3)$$

Maximizing $m$ in the new, massively parallel submarine cable context asks for revisiting the following fundamental questions:

i) Do we need gain-flattening filters (GFFs) in conjunction with optical amplifiers? GFFs are universally used in all submarine systems today but are lossy optical elements and hence attenuate parts of the signal spectrum, thereby reducing the signal power after amplification, which wastes precious cable supply power and potentially reduces $m$.

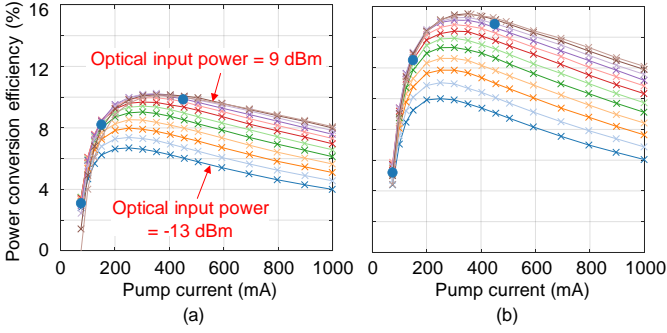

Fig. 2. Power conversion efficiency of the EDFAs used in this experiment, measured (a) with and (b) without GFF, as a function of the pump current for various optical input powers from -13 dBm to 9 dBm in 2-dB increments. The six solid blue circles are the operating points used in our experiments, cf. Table I.

TABLE I
POWER CONVERSION EFFICIENCY AT THE CHOSEN OPERATING POINTS

| Pump current | With GFF | Without GFF |
|---|---|---|
| 75 mA | 3.1 % | 5.2 % |
| 150 mA | 8.2 % | 12.5 % |
| 450 mA | 9.9 % | 14.9 % |

representing signal plus noise power at frequency bin $k$) to the ASE power $N_k$ generated by in-line amplifiers (square marker in the inset), interpolated between two empty slots (cross markers in the inset), i.e., $SNR_k = S_k/N_k$.

The emulated WDM channels at the WSS output are boosted by a TX EDFA and are attenuated by a variable optical attenuator (VOA) to produce a set of desired optical launch powers $\boldsymbol{P}_{1:40}$; we use the notation $\boldsymbol{X}_{1:K} \coloneqq [X_1, \ldots, X_K]$ throughout the paper. An example flat TX signal power allocation across a system bandwidth of 4 THz is shown in Fig. 1(b), with a total launch power of 13 dBm. The line system comprises 12 spans of 62-km Corning® Vascade® EX3000 fiber with 0.16-dB/km loss. Each span is padded by a VOA in order to realize a span loss of ~16.5 dB and to operate our 744-km straight-line system in a lower-OSNR regime pertinent to the targeted massively parallel submarine application [1]–[5]. Since launch powers are low and fiber nonlinearities are negligible (as quantitatively verified below), padding at the beginning of a span is equally permissible as padding at the end. Each span is followed by a custom-designed single-stage EDFA with a removable GFF, as we want to compare the capacity gain by removing GFFs in an otherwise identical system. An example RX PSD for the flat TX signal power allocation in Fig. 1(b) is shown in Fig. 1(c), when all the GFFs are removed from the link and the EDFA pump currents are set to 150 mA such that a total optical power of 13 dBm is obtained after each EDFA.

The custom EDFAs used in this experiment were fabricated by Amonics, with a single-stage 980-nm co-pumped design. The operating output power was designed to be varied between 13 dBm and 19 dBm. The noise figure ranges from 4.5 dB to 5 dB, depending on the operating point. The GFF is connectorized to the output of the EDFA such that a comparison between systems with and without GFFs can readily be made.

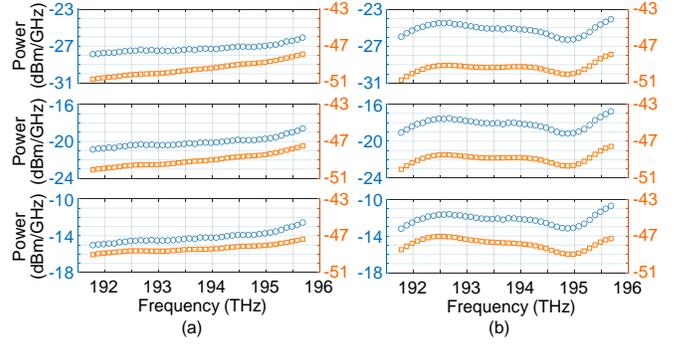

Fig. 3. Signal powers (blue circles, left $y$-axes) and noise powers (orange squares, right $y$-axes) measured at the output of a single EDFA when the input signal powers are flat; (a) with GFF and (b) without GFF, using the pump currents of 75 mA (top), 150 mA (middle), and 450 mA (bottom), producing $\mathcal{P}_O$ = 6.2, 13.0, 19.0 dBm, respectively.

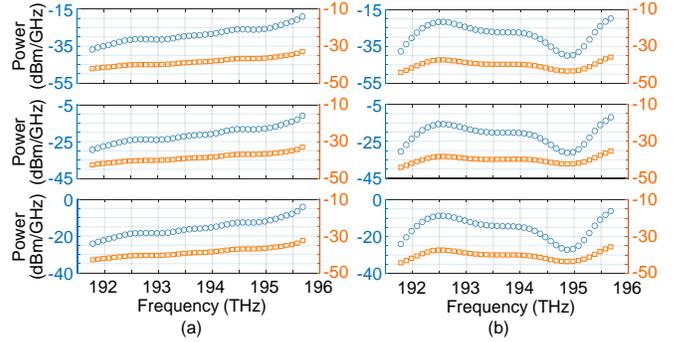

Fig. 4. Signal powers (blue circles, left $y$-axes) and noise powers (orange squares, right $y$-axes) measured after 12-span transmission of flat TX signal powers; (a) with GFFs and (b) without GFFs in the line system, using the pump currents of 75 mA (top), 150 mA (middle), and 450 mA (bottom), producing $\mathcal{P}_O$ = 6.2, 13.0, 19.0 dBm, respectively.

The current-voltage curves of the pumps allow us to estimate the electrical pump power for each chosen pump current.

From a component perspective, the electrical-to-optical power conversion efficiency (PCE) [25, Ch. 8] of the EDFA is one of the most crucial factors to maximize submarine cable capacity under an electrical supply power constraint. The PCE is defined as

$$PCE = \frac{\mathcal{P}_O - \mathcal{P}_I}{\mathcal{P}_E}, \quad (5)$$

where $\mathcal{P}_I$, $\mathcal{P}_O$, and $\mathcal{P}_E$ denote the optical input power to the EDFA, the optical output power (measured *after* an optional GFF attached to the output of the EDFA), and the electrical pump power, respectively. Figure 2 shows the PCE of our Amonics EDFAs as a function of the pump current, for various optical input powers $\mathcal{P}_I$ ranging from -13 dBm to 9 dBm in 2-dB increments. The three solid circles at the pump currents of 75, 150, 450 mA, respectively for the cases with and without GFFs in Figs. 2(a) and 2(b), correspond to pump powers of $\mathcal{P}_E$ = 100, 207, 685 mW and are the operating points chosen for our experiment. In the 12-span link, these operating points consume a total electric pump power (across all line amplifiers) of $\mathcal{P}_E$ = 1.09, 2.27, 7.53 W and produce a total optical output power per amplifier of $\mathcal{P}_O$ = 6.2, 13.0, 19.0 dBm (measured after the GFF), all with a gain (i.e., $(\mathcal{P}_O - \mathcal{P}_I)/\mathcal{P}_I$) of ~16.2 dB with GFFs and of ~14.1 dB without GFFs. A lower gain is observed when the GFFs are removed from the system, since the EDFAs operate in their saturation regimes. Table I shows the measurement data for a total of 6 system scenarios discussed in this paper. Figure 3 shows signal and noise power profiles of a *single* EDFA at the three operating points, with and without GFF. When the input signal powers are flat across the studied 4-THz amplification band, a gain tilt of up to 2.5 dB is observed at the output of the EDFA with a GFF (cf. left figures in Fig. 3), which is a 1-dB larger tilt than what is found for the nominal operating condition for which the amplifiers were designed. A deviation from the nominal amplifier operating condition was necessary due to the need to reduce the OSNR through loss-padding in our experimental link. When the GFF is removed (cf. right figures in Fig. 3), the largest signal power excursion at the EDFA output is similar to the case with the GFF, but this excursion is caused by a gain *ripple* instead of a gain *tilt*. It is clear from Figs. 2 and 3 that the GFF wastes a significant amount of electrical supply power to flatten the EDFA's spectral gain profile. After 12-span transmission of a flat TX signal power distribution, similar RX power profiles as those of the back-to-back measurements are observed, as shown in Fig. 4, both with and without the GFFs in the line system, yet with much increased signal power excursions.

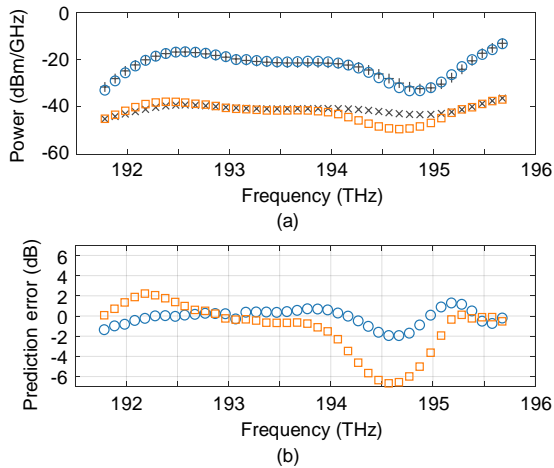

Fig. 5. Prediction by a physics-based simulation: (a) Measured RX signal (black pluses) and noise (black crosses) powers after 12-span transmission of a flat TX signal power distribution, and the predicted RX signal (blue circles) and noise (orange squares) powers, (b) the prediction errors for the RX signal (blue circles) and noise (orange squares) powers.

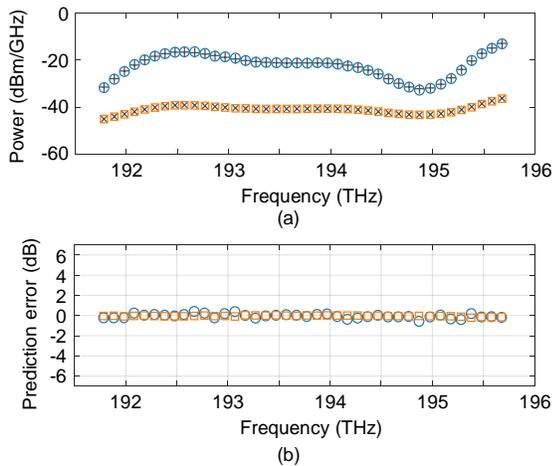

Fig. 6. Prediction by an NN: (a) Measured RX signal (black pluses) and noise (black crosses) powers after 12-span transmission of a flat TX signal power distribution, and the predicted RX signal (blue circles) and noise (orange squares) powers, (b) the prediction errors for the RX signal (blue circles) and noise (orange squares) powers.

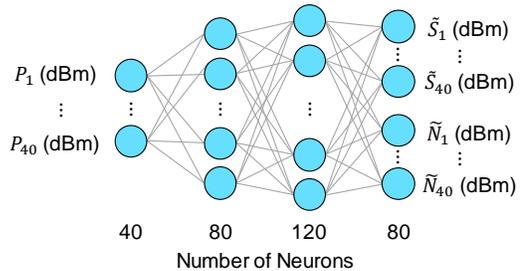

Fig. 7. Structure of the NN.

In order to see how far physics-based EDFA simulations can correctly predict the RX PSD from a TX PSD, we use the OASIX simulation software [26], considered to be one of the most accurate simulation tools for EDFAs, with complete knowledge of type and length of the erbium-doped fiber used in the amplifiers as well as any input and output component losses within the EDFA. Spectral hole burning is also taken into account in the simulation. In order to get somewhat reliable results, it is necessary to use the exact manufacturing lot data of the specific erbium-doped fiber. For the test case of $\mathcal{P}_E = 2.27$ W without GFFs in the 12-span link (this will be used as our test case for illustration throughout the paper), Fig. 5(a) shows the comparison of the RX signal and noise powers between experimental and simulated results, when TX signal powers are flat. The corresponding prediction errors $S_k/\tilde{S}_k$ and $N_k/\tilde{N}_k$ in dB, for $k = 1, \dots, 40$, are shown in Fig. 5(b), where $S_k$ and $N_k$ denote the measured signal and noise powers, respectively, and $\tilde{S}_k$ and $\tilde{N}_k$ are the values obtained by simulation. The mean and maximum absolute errors of the RX signal and noise powers between measurement and simulation amount to 1.25 dB and 6.67 dB, respectively. This large discrepancy motivates us to use an NN instead of a physics-based model, as described in the following section, which produces mean and maximum absolute errors of only 0.11 dB and 0.57 dB, as shown in Fig. 6.

### III. TRAINING THE NN WITH EXPERIMENTAL DATA

Since each of the signal and noise powers ($\boldsymbol{S}_{1:40}$, $\boldsymbol{N}_{1:40}$) at the output of the link depends on the full set of launch powers $\boldsymbol{P}_{1:40}$ in a way that is difficult to accurately model based on amplifier physics as discussed in Sec. II, we resort to machine learning and construct an NN as a digital twin of our experimental link. As shown in Fig. 7, we chose an NN with 40 input neurons ($\boldsymbol{P}_{1:40}$), two hidden layers with 80 and 120 neurons each, and 80 output neurons for the predicted signal and noise powers ($\tilde{\boldsymbol{S}}_{1:40}$, $\tilde{\boldsymbol{N}}_{1:40}$) at the output of the link, where $P_k$, $\tilde{S}_k$, and $\tilde{N}_k$ are all expressed in log scale since *(i)* linear-scale powers cause numerical problems due to a wide dynamic range and a high resolution, and *(ii)* the log-scale powers are more closely related to the capacity as given by Eq. (4) than the linear-scale powers. Linear, sigmoid, and softplus activation functions [27] are used. Numbers of neurons and activation functions are chosen to minimize the mean absolute error (MAE) between measurement $(S_{1:40}, N_{1:40})$ and prediction $(\tilde{S}_{1:40}, \tilde{N}_{40})$, defined as

$$MAE := \frac{1}{80} \sum_{k=1}^{40} \left( |10 \log_{10} S_k - 10 \log_{10} \tilde{S}_k| + |10 \log_{10} N_k - 10 \log_{10} \tilde{N}_k| \right). \quad (6)$$

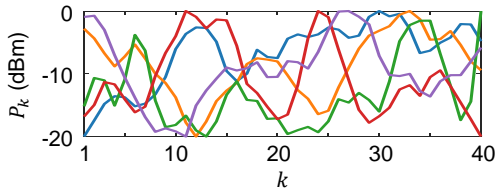

Fig. 8. Five exemplary random TX power profiles $P_{1:40}$ with $\mathcal{F} = 20$ dB.

Note that the input to our NN assumes a noise-free signal, while in practice the TX SNR is limited to ~45 dB due to the finite extinction of the WSS, leading to non-zero TX noise powers, cf. Fig. 1(b). However, this has minimal impact on the performance of the NN, since the optical line SNR is substantially lower than the TX SNR; e.g., a 45-dB TX SNR reduces a 20-dB RX SNR only by 0.014 dB, cf. Eq. (2).

Each NN training process starts by configuring one of two link setups (i.e., with and without GFFs) and choosing one of three total available electrical supply power levels $\mathcal{P}_E$, for a total of 6 different link operating conditions, cf. Table I. We consider only *electrical pump powers* and ignore less fundamental overheads from amplifier control [28]. The overall electrical pump power is spread approximately evenly across the 11 in-line EDFAs such that the optical output power summed over all 40 signal channels ($\mathcal{P}_O = \sum_{k=1}^{40} P_k$) is the same for all EDFAs. The TX VOA is adjusted to provide the same total TX power $\mathcal{P}_O$ during this process. The whole configuration process is automated and controlled by software. Next, we measure $S_{1:40}$ and $N_{1:40}$ for 1440 randomly generated sets $P_{1:40}$ subject to maintaining $\sum_{k=1}^{40} P_k = \mathcal{P}_O$. Each randomly generated set of powers has a distinct peak-to-peak channel power excursion of $\mathcal{F} = \max_{i,j}(|P_i - P_j|)$ that we gradually increase from 6 dB to 45 dB; 5 representative instances of TX signal power allocations with 20-dB peak-to-peak excursion are depicted in Fig. 8, cf. Appendix B.

We then use the Adam algorithm [29] to train the NN, which has recently been adopted widely in many deep learning applications for its excellent performance, with L2 regularization [30] to avoid potential overfitting that can be caused by a small training data set. After proper training, the NN acts as a digital twin of the true optical fiber link and *predicts* output signal and noise powers $(\tilde{S}_{1:40}, \tilde{N}_{1:40})$ from any input signal power allocation $P_{1:40}$. For each of the RX power profiles, the accuracy of the NN prediction can be evaluated by $MAE$, cf. Eq. (6). Figure 9(a) shows mean (circles) and standard deviation (crosses) of the 1440 $MAE$ values for various training set sizes ranging from 100 to 1400 for the running test case of $\mathcal{P}_E = 2.27$ W without GFFs. The temporal convergence behavior of the NN for a training set size of 1300 is shown in Fig. 9(b), when min-max normalization [31] is used to set the range in $[0, 1]$; note that $MAE$ in Fig. 9(b) is with respect to the min-max normalized log-scale $S_{1:40}$ and $N_{1:40}$ calculated during the NN training process, whereas $MAE$ in Fig. 9(a) is with respect to unnormalized log-scale $S_{1:40}$ and $N_{1:40}$ showing the performance of the fully-trained NNs.

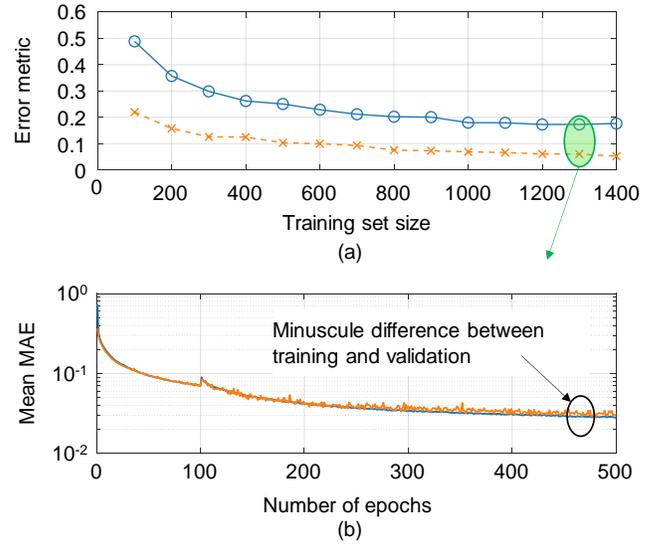

Fig. 9. (a) Mean (circles) and standard deviation (crosses) of 1440 $MAE$ values for a varying training set size; (b) convergence of the NN without overfitting for a training set size of 1300.

In this paper, we use the MAE for training and evaluation of the NN, since it makes the training more robust to outliers [32]. In our experimental system, when the TX signal powers are greatly varied across the channels, the measured RX signal powers on some channels occasionally become very small (e.g., $< -30$ dB), leading to large squared errors between the measurement and the prediction. The MAE does not have this problem, but its derivative used in training of the NN is not defined at zero. We can circumvent this problem by, e.g., assigning a random value $\in \{\pm 1\}$ for the derivative of zero MAE. The impact of this mathematical imperfection on training is trivial since the probability of occurrence of zero MAE anywhere inside the NN is negligibly small.

Following conventional NN nomenclature, training is performed in multiple *batches* (i.e. subsets of the whole training set) per *epoch* (i.e. iteration of the learning algorithm for the whole training set). The batch size for training is reduced at the 100th epoch for finer convergence [33], [34], which produces the bump in Fig. 9(b). The accuracy of the NN represented by the two error metrics shown in Fig. 9(a), i.e., mean and standard deviation of $MAE$, generally improves as the training set size increases, but the improvement saturates at training set sizes beyond ~1000. We consequently chose a training set size of 1300 that produces 0.18 dB of mean $MAE$; the remaining 140 of the overall 1440 measurement data are used for validation of the NN. For the chosen training (blue) and validation (orange) sets, the NN converges rapidly with the number of epochs, as shown in Fig. 9(b). The minuscule difference between the mean $MAE$ values of the training and validation sets at a large number of epochs indicates that overfitting is unlikely [35], i.e., the numbers of neurons and NN layers are chosen adequately for the dataset. We also confirmed by Monte-Carlo cross-validation with random sub-sampling [36] that the chosen NN topology is free from the overfitting or the selection bias problems. The convergence behavior of the NN is similar for all the 6 test cases of Table I, and for no test case does the mean $MAE$ exceed 0.31 dB.

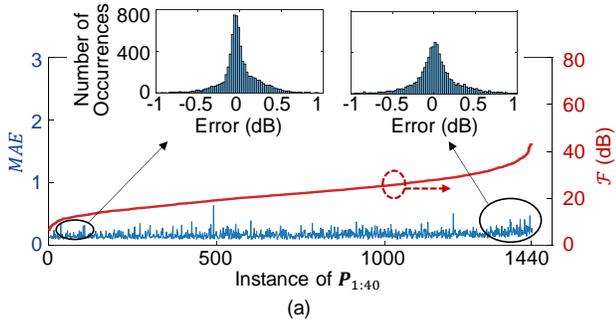

(a)

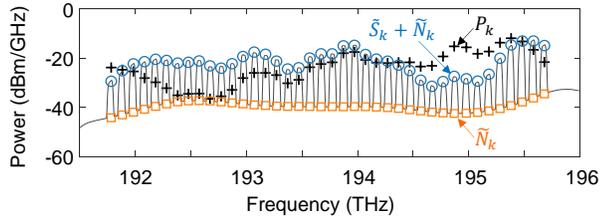

(b)

Fig. 10. (a) Accuracy of SNR prediction of the NN, and (b) an example of TX powers (black pluses), measured RX PSDs (black solid line), and NN-predicted signal+noise (blue circles) and noise (orange squares)

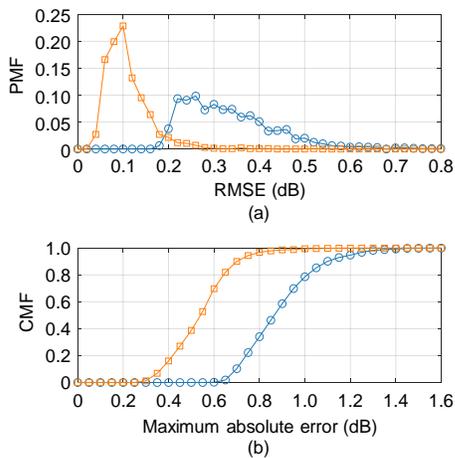

Fig. 11. (a) PMF of the RMSEs, and (b) CMF of the maximum absolute errors for the signal (blue circles) and noise (orange squares) powers.

Figure 10(a) shows the $MAE$ values for 1440 test cases (left $y$-axis) with $\mathcal{F}$ increasing from 6 dB to 45 dB (right $y$-axis). For all test cases across this very wide range of $\mathcal{F}$, the NN predicts the RX signal and noise powers with very small prediction errors. The small prediction errors justify the optimization of signal power allocations based on NNs. A representative example of $\boldsymbol{P}_{1:40}$ (black pluses), the corresponding measured RX PSDs (black solid line), and the NN-predicted $\widetilde{\boldsymbol{S}}_{1:40} + \widetilde{\boldsymbol{N}}_{1:40}$, $\widetilde{\boldsymbol{N}}_{1:40}$ (blue circles, orange squares) are depicted in Fig. 10(b), with great agreement between measurement and prediction. In addition, Figs. 11(a) and 11(b) show, respectively, the probability mass function (PMF) of the root mean-squared errors (RMSEs) $\frac{1}{40}\sum_{k=1}^{40}|10\log_{10} S_k - 10\log_{10} \tilde{S}_k|^2$ and $\frac{1}{40}\sum_{k=1}^{40}|10\log_{10} N_k - 10\log_{10} \widetilde{N}_k|^2$, and the cumulative mass function (CMF) of the maximum absolute errors $\max_k |10\log_{10} S_k - 10\log_{10} \tilde{S}_k|$ and $\max_k |10\log_{10} N_k - 10\log_{10} \widetilde{N}_k|$ obtained over the 1440 random power profiles.

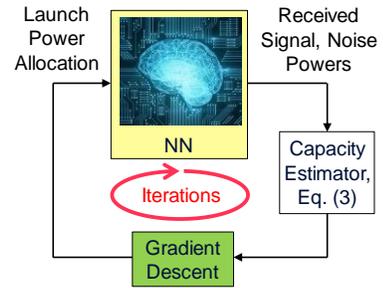

Fig. 12. Capacity optimization using NN and GD.

## IV. CAPACITY MAXIMIZATION AND VERIFICATION

We next perform gradient descent (GD) capacity maximization off-line, based on the trained NN, cf. Fig. 12 and Appendix C. We perform GD iterations until there is no more noticeable increase in $C$. The result is a capacity-maximizing TX power profile $\boldsymbol{P}_{1:40}$. Figure 13(a) shows, for the running test case of $\mathcal{P}_E = 2.27$ W without GFFs, three example capacity optimizations, one starting from a flat $\boldsymbol{P}_{1:40}$ (blue) and the other two from initial conditions with poorer capacity. All 3 initial conditions converge to the same maximum capacity. Even more impressively, Fig. 13(b) shows initial (blue crosses) and converged (orange dots) capacities for *all* of the 1440 randomly chosen power profiles with varying $\mathcal{F}$ (red dots). All but 18 initial conditions converge to the same maximum capacity, even for peak channel power deviations as much as 45 dB! Figures 13(c) shows the random TX power profiles (top), corresponding RX power profiles (middle), and the RX SNRs (bottom) across the system bandwidth, at iteration 0 (left) and after convergence (right) of the GD optimization, where 18 outlier profiles are removed from the right figures. The mis-convergence for these outliers is attributed to the smoothening operation that intervenes after every GD iteration, cf. Appendix C. Remarkably, the converged SNR distribution is reasonably flat to within 4 dB in most cases, cf. inset to the bottom of Fig. 13(c), with minimal capacity variations between these converged solutions, cf. Fig. 13(b). The capacity of a completely flat SNR is 25.6 Tb/s, which is close to the experimental optimum of 25.9 Tb/s but further off the capacities of a flat TX signal power profile (24.8 Tb/s) and a flat RX signal power profile (24.5 Tb/s), in contrast to the findings of Ref. [37] for conventional systems using GFFs. The convergence of almost all initial TX power profiles to the same capacity indicates that the capacity surface is *nearly concave* within the boundaries of our experimental conditions, hence there may be *only one optimal* TX power allocation. The small spread in the converged RX SNRs in Fig. 13(c) implies that the capacity surface is almost flat near the peak so that small deviations from the optimal TX power allocation only leads to an insignificant capacity loss, which causes the GD algorithm to terminate before reaching the absolute peak.

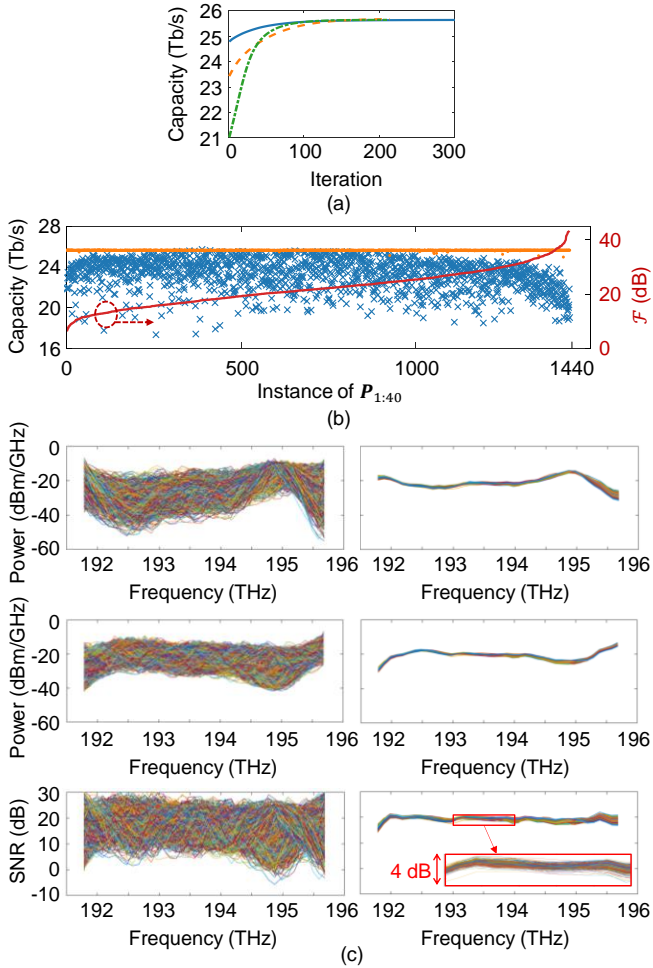

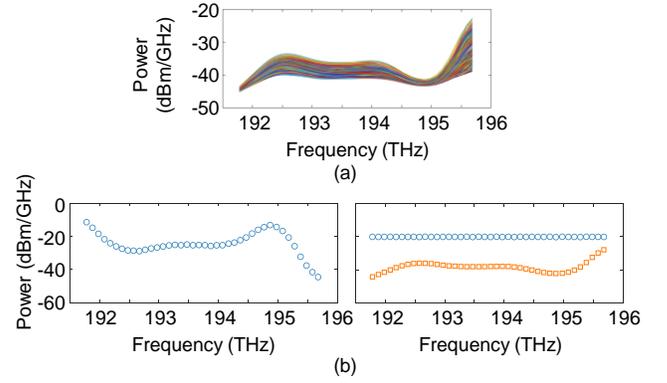

Fig. 13. (a) Typical temporal convergence behavior of the GD optimization, (b) capacity of the initial (blue crosses) and converged (orange dots) TX power profiles for all 1440 random initial TX power profiles, (c) 1440 random TX power profiles (top), corresponding RX signal+noise powers (middle), and RX SNRs (bottom), each at iteration 0 (left) and after convergence (right) of the GD optimization operated on the NN.

Fig. 14. (a) RX noise powers for the 1440 random TX power profiles, and (b) the TX powers optimized using the waterfilling strategy (left) and the corresponding RX signal+noise (right, blue circles) and noise (right, orange squares) powers.

Importantly, our capacity maximization approach does *not* subjectively favor any classical TX power allocation strategy based on possibly misguiding intuition (e.g., "flat RX signal power profile", "flat TX signal power profile", "flat SNR", or "waterfilling"), but objectively optimizes the TX signal powers solely by following the gradient trajectory of ascending system capacity. Also, note that an *experimental* GD solution is uniquely enabled by our NN approach for the following two reasons: *(i)* First, as shown in Appendix C, estimating only a single gradient requires 41 measurements of 4-THz RX PSDs in our case. In our fully automated system, we are able to measure 180 RX PSDs per hour, hence it would require >11 years to perform a full GD optimization for 1440 TX power profiles with 300 GD iterations! On the other hand, the optimization process takes only 9 hours using the NN approach. This >10,000× speed-up reveals the power of machine learning, enabling us to experimentally determine a multi-dimensional capacity surface, which would have been impossible by physical experiments alone. To the best of our knowledge, this is the first experimental observation of the optical WDM channel capacity surface as a joint function of the constituent channel powers. *(ii)* Second, a small change of the TX power on a certain channel, as described in Eq. (9) of Appendix C by $\varepsilon_k$, would produce such a small change of the capacity in experiments that can easily be buried in measurement errors of the signal and noise powers. Reducing measurement errors requires averaging over multiple OSA sweeps and/or high-resolution OSA sweeps, both of which substantially increase the measurement time.

In order to compare our results to the standard waterfilling solution [10], [11], we first measure the spectral noise distribution for all 1440 TX signal power profiles ($\mathcal{P}_E = 2.27$ W without GFFs), cf. Fig. 14(a). While the standard waterfilling framework assumes that the spectral noise distribution remains the same when re-allocating signal powers across the spectrum, Fig. 14(a) clearly shows that the TX power distribution for a fixed total launch power can change the RX noise powers in an optical fiber link, by as much as 15 dB! This invalidates the assumption on signal-independent noise for the well-known waterfilling algorithm to be optimal. As the noise powers vary with signal powers, we cannot obtain an analytical solution to the waterfilling strategy but resort to an iterative method for a numerical solution. Namely, given initial TX powers and the observed RX signal+noise powers, we slightly increase those TX powers that produce lower RX signal+noise powers than a flat water level, and vice versa. The amount of the change in TX powers is propotional to the distance of the RX signal+noise powers from the flat water level. By repeating this procedure, the RX signal+noise powers converge to a flat water level. When the iterative waterfilling strategy is applied to the NN for the test case of $\mathcal{P}_E = 2.27$ W without GFFs, all the 1440 TX power profiles converge to a single TX power profile shown in Fig. 14(b). For the cases of $\mathcal{P}_E = 1.09, 2.27, 7.53$ W without GFFs, the capacities obtained from the iterative waterfilling are smaller than those obtained by the GD by 6.0%, 6.2%, 13.0%, respectively.

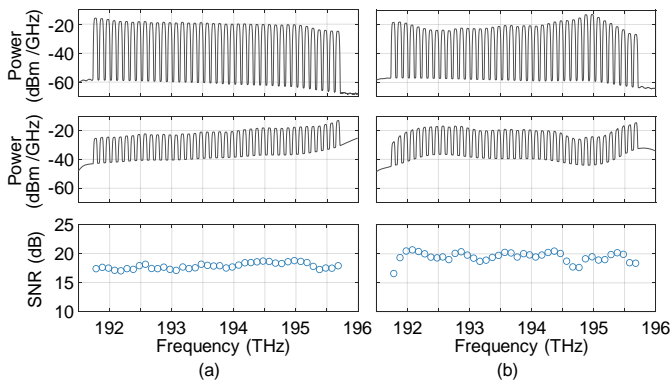

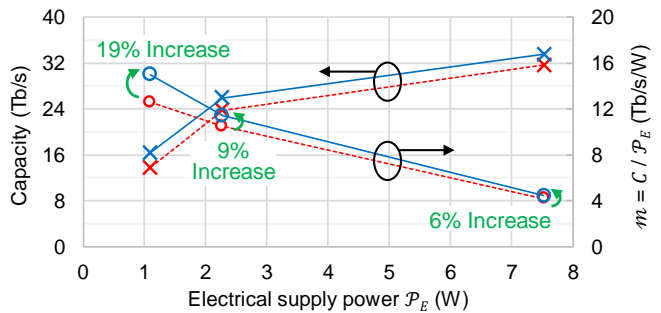

Fig. 15. TX PSDs (top), RX PSDs (middle), and SNRs (bottom) measured in the experiment, when the GD-NN-optimized TX signal powers are transmitted over 12 spans, (a) with and (b) without the GFFs in the line system, using the pump current of 150 mA.

Fig. 16. Optimized capacity $C$ (crosses) and power efficiency $m$ (circles) as a function of the total electric pump power in systems with (dashed) and without (solid) GFFs. The capacity gains are estimated for the most conservative scenario with $\eta = 1$, and are increased with $\eta < 1$.

As a last step, we validate the results of the NN-based GD optimization by loading the optimized TX power profiles into the experimental WDM emulator and measuring the resulting RX power profile. The TX, RX PSDs and SNRs as measured by experiment for a pump current of 150 mA per amplifier are shown in Fig. 15. The capacity predicted by the NN is within a 1.1% error of the experimentally measured capacity in all the test cases of Table I. We also verify that the maximum channel power occurring anywhere within the system for the optimized power profile is below -4 dBm per 50-GHz slot for $\mathcal{P}_E = 1.09$ W, both with and without GFFs. This justifies the initial assumption of neglecting fiber Kerr nonlinearities (For reference, the nonlinearity-optimized power used in a deployed 5500-km submarine cable is 3 dBm per 50-GHz channel [24].)

Figure 16 shows the actually measured capacity $C$ (left axis) and the power efficiency figure of merit $m$ (right axis). Dashed lines represent systems with GFFs and solid lines without GFFs, all with optimized TX power allocations. The experimental results show that: *(i)* systems without GFFs achieve a better power efficiency than systems with GFFs, *(ii)* $m$ increases with decreasing $\mathcal{P}_E$ until the EDFA pump current approaches the pump's lasing threshold, even at a significantly reduced EDFA power conversion efficiency of only 3.1% at that operating point (cf. Fig. 2); hence, operating the pumps at higher power (and hence at higher efficiencies) and sharing their power across multiple EDFAs will further increase $m$; and *(iii)* when the system operates at maximum efficiency (at largest $m$), both $C$ and $m$ can be increased by 19% by eliminating GFFs from the system (with $\eta = 1$, assuming zero implementation penalty in the capacity estimate of Eq. (4)). For $\eta = 0.5$ and $\eta = 0.25$, which correspond to implementation penalties of 3 dB and 6 dB in SNR, respectively, the GD-NN approach predicts greater gains of up to 23% and 29% in $C$ and $m$ from eliminating GFFs.

## V. CONCLUSION

We used experimental signal and noise data from a 12-span 744-km straight-line EDFA link to train an NN as a digital twin of the experimental system. The NN accurately predicts received signal and noise powers for arbitrary transmit signal power distributions, even without GFFs as part of the link. A gradient descent based transmit power profile optimization performed on the NN is about 10000 times faster than what would be possible using measurements alone and objectively predicts optimized launch power profiles. In the context of a massively parallel fiber system under a total electrical supply power constraint across the fibers (such as a cost-optimized submarine optical cable), we demonstrate substantial improvements in achievable cable capacity.

We did not load signal power on half of the total WDM bandwidth in the experimental system, in order to measure the SNR for each channel using an OSA. We also relied on an assumed value for the transponder implementation penalty and ignored NL impairments, when estimating the system capacity from the SNRs. However, in deployed systems with real-time transponders, the proposed NN-based capacity maximization can potentially produce more accurate results, since: *(i)* the capacity of each WDM channel can be obtained from the true SNR that the transponders estimate from the recovered signal constellation, and *(ii)* the whole frequency band can fully be populated.

## APPENDIX

### A. *The Small Benefit of Waterfilling at High SNR*

Spectral waterfilling is widely used in wireless communications to maximize the *aggregate* capacity of a channel with frequency-dependent SNR [10]–[13] by optimizing the spectral TX power profile relative to flat TX power allocation. In contrast, long-haul optical transmission systems have historically been designed to provide the same, fair interface rates for all WDM channels, i.e., to provide flat SNRs across the system bandwidth [24]. In order to see the potential benefit that optimal TX power allocation may bring relative to this baseline in a WDM system, we simplistically assume (only within this appendix) that there is no dependence between noise powers and signal powers across WDM channels; the rest of this paper takes all such dependencies into account through NN-based channel power optimization. Within this appendix, we assume 40 WDM channels with frequency-dependent noise that has a flat spectrum in each channel's frequency band, characterized by *(i)* the ratio $\mathcal{F}_N$ of the maximum spectral noise power to the minimum spectral noise power at the receiver, *(ii)* a logarithmically uniform spectral noise power distribution across the channels, and *(iii)* an overall SNR (denoted by $\overline{SNR}$) calculated as the total RX signal power summed over all channels to the total RX noise power summed over all channels. Under these assumptions, Fig. 17 shows the ratio of the achievable aggregate system capacity using waterfilling [10]–[13] to the achievable aggregate system capacity of the flat-SNR baseline. In this simplistic scenario, the capacity gain from waterfilling amounts only to $< 10\%$ for typical long-haul optical transmission systems, even when removing GFFs from the system ($\mathcal{F}_N \lesssim 12.2\ dB$, $\overline{SNR} \gtrsim 12.4\ dB$).

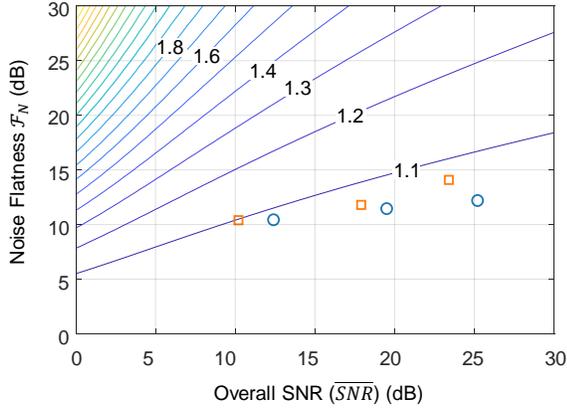

Fig. 17. Capacity gain using waterfilling relative to the constant-SNR power allocation across the system bandwidth. The experimental conditions are indicated by markers, for the test cases with (orange squares) and without (blue circles) GFFs.

### B. Generation of Random TX Power Profiles

We first create $P_k$ for $k = 1, \ldots, 40$ by cumulative random walks; more specifically, $P_k = P_{k-1} + U_k$, where $P_0 \coloneqq 0$ and $U_k$ is randomly drawn from the unit interval $(-0.5, 0.5)$. Then, implausibly fast changes of $P_k$ over a narrow frequency range is avoided by applying a weighted moving average over 3 frequency bins to each TX power profile. We also ensure that the 1440 random power profiles do not have similar shapes by maximizing the relative entropies between the power profiles normalized to sum to one. More specifically, assuming that $n - 1$ random power profiles $\boldsymbol{P}_{1:40}^{(i)}$ for $i = 1, \ldots, n-1$ have already been created, the $n$-th profile $\boldsymbol{P}_{1:40}^{(n)}$ is generated by first creating a set of 100 random candidate profiles $\boldsymbol{Q}_{1:40}$, then picking the one that has the greatest minimum relative entropy with all the $n-1$ previously generated profiles, i.e., we choose

$$\boldsymbol{p}_{1:40}^{(n)} = \underset{\boldsymbol{q}_{1:40} \in \mathcal{Q}}{\text{argmax}} \left[ \underset{i \in [1, n-1]}{\min} D\big(\boldsymbol{p}_{1:40}^{(i)} \parallel \boldsymbol{q}_{1:40}\big) \right], \quad (7)$$

where $\boldsymbol{p}_{1:40}^{(i)}$ and $\boldsymbol{q}_{1:40}$ are the values $\log_{10} \boldsymbol{P}_{1:40}^{(i)}$ and $\log_{10} \boldsymbol{Q}_{1:40}$, linearly transformed such that they vary between 0.1 and 1; $\mathcal{Q}$ is the set of such generated 100 candidates $\boldsymbol{q}_{1:40}$ and $D(\cdot \parallel \cdot)$ denotes the relative entropy. This ensures that the new profile is not too close to any of the formerly generated profiles in terms of its relative entropy. We repeat this process from $n = 1$ until the required number of random power profiles are obtained and re-scale them to their target $\mathcal{F}$.

### C. Approximate Stochastic Gradient Descent Method

Denoting the true power transfer function of the optical link by $h: \boldsymbol{P}_{1:40} \to (\boldsymbol{S}_{1:40}, \boldsymbol{N}_{1:40})$, and by knowing that the capacity as given in Eq. (4) is a function of the RX powers $(\boldsymbol{S}_{1:40}, \boldsymbol{N}_{1:40})$, the gradient of the capacity $C$ with respect to the TX powers $\boldsymbol{P}_{1:40}$ can be written as

$$\nabla_{\boldsymbol{P}_{1:40}} C(h(\boldsymbol{P}_{1:40})) = \left[ \frac{\partial C}{\partial P_1}, \ldots, \frac{\partial C}{\partial P_{40}} \right]. \quad (8)$$

Since the NN acts as an approximate function $\tilde{h}: \boldsymbol{P}_{1:40} \to (\tilde{\boldsymbol{S}}_{1:40}, \tilde{\boldsymbol{N}}_{1:40})$ to the true function $h$, the partial derivatives of Eq. (8) can be computed approximately using the NN ($\tilde{h}$) as

$$\frac{\partial C}{\partial P_k} \approx \frac{C\left(\tilde{h}(\delta(\boldsymbol{P}_{1:40} + \boldsymbol{\varepsilon}_k))\right) - C\left(\tilde{h}(\boldsymbol{P}_{1:40})\right)}{\varepsilon} \quad (9)$$

for $k = 1, \ldots, 40$, where $\varepsilon$ is a small positive number and $\boldsymbol{\varepsilon}_k \coloneqq [0, \ldots, \varepsilon, \ldots, 0]$ has the only non-zero element $\varepsilon$ at the $k$-th position, and $\delta$ is a scaling factor to fulfill the total power constraint $\sum_{k=1}^{40} P_k = \mathcal{P}_O$. Namely, the partial derivative in Eq. (9) quantifies how much the *sum capacity* over all 40 channels increases if the TX power is increased *only* on the $k$-th channel by $\varepsilon$ (followed by the scaling with $\delta$). Note that estimation of the gradient in Eq. (8) at a point $\boldsymbol{P}_{1:40}$ using the partial derivatives in Eq. (9) requires evaluation of $\tilde{h}$ with 41 different TX power profiles (i.e., $\boldsymbol{P}_{1:40}$ and $\boldsymbol{P}_{1:40} + \boldsymbol{\varepsilon}_k$ for $k = 1, \ldots, 40$). This is an important reason for not being able to optimize the TX power profile solely based on experiments, as addressed in Sec. IV. Using the gradient, a typical stochastic GD algorithm iteratively updates the TX power profile as

$$\boldsymbol{P}_{1:40} \leftarrow \boldsymbol{P}_{1:40} + \mu \nabla_{\boldsymbol{P}_{1:40}} C(\tilde{h}(\boldsymbol{P}_{1:40})), \quad (10)$$

with $\mu$ being a learning rate. In this work, since the NN is trained with only smooth TX power profiles, a weighted moving average is performed on $\boldsymbol{P}_{1:40}$ after every GD iteration, to prevent the updated NN inputs from diverging out of the statistical boundaries of the training set. Specifically, we smoothen the TX power profile as $P_k \leftarrow 0.03 P_{k-1} + P_k + 0.03 P_{k+1}$, with the weighted moving average window properly truncated at the ends of the WDM channels. The updated TX power profile is then scaled such that $\sum_{k=1}^{40} P_k = \mathcal{P}_O$.

ACKNOWLEDGMENT

We acknowledge Amonics Corp for designing and supplying the EDFAs with removable GFFs, and Corning for the loan of the EX3000 fiber used in this experiment.